\documentclass[preprint,showpacs,aps,preprintnumbers,amsmath,amssymb]{revtex4-1}

\usepackage{graphicx,epsfig}% Include figure files
\usepackage{dcolumn}% Align table columns on decimal point
\usepackage{bm}% bold math
\newcommand{\beq}{\begin{equation}}
\newcommand{\eeq}{\end{equation}}
\newcommand{\beqa}{\begin{eqnarray}}
\newcommand{\eeqa}{\end{eqnarray}}
\newcommand{\lslash}[1]{#1\llap/}

\newcommand{\calE}{{\cal E}}

\newcommand{\Eq}[1]{Eq.\ (\ref{#1})}
\newcommand{\Eqs}[2]{Eqs.\ (\ref{#1}) and (\ref{#2})}
\newcommand{\Eqsto}[2]{Eqs.\ (\ref{#1})-(\ref{#2})}
\newcommand{\Eqss}[3]{Eqs.\ (\ref{#1}), (\ref{#2}) and (\ref{#3})}

\newcommand{\Section}[1]{Section\ \ref{#1}}
\newcommand{\bse}{\begin{subequations}}
\newcommand{\ese}{\end{subequations}}
\newcommand{\bpmat}{\begin{pmatrix}}
\newcommand{\epmat}{\end{pmatrix}}
\newcommand{\bwt}{\begin{widetext}}
\newcommand{\ewt}{\end{widetext}}

\begin{document}

%\preprint{ICN/000-HEP}

\title{Neutrino dispersion relation in a  magnetized
multi-stream matter background}
% Force line breaks with \\
\author{Jos\'e F. Nieves$^{a}$}
\email{nieves@ltp.uprrp.edu}
\author{Yaithd D. Olivas$^{b}$}
\email{dolivas@live.com.mx}
\author{Sarira Sahu$^{b,c}$ }
\email{sarira@nucleares.unam.mx}

\affiliation{$^{a}$Laboratory of Theoretical Physics,
  Department of Physics, University of Puerto Rico 
  R\'{\i}o Piedras, Puerto Rico 00936}
\affiliation{$^{b}$Instituto de Ciencias Nucleares, Universidad Nacional Aut\'onoma de M\'exico, 
Circuito Exterior, C.U., A. Postal 70-543, 04510 Mexico DF, Mexico}
\affiliation{$^{c}$Astrophysical Big Bang Laboratory, RIKEN, Hirosawa, Wako, Saitama 351-0198, Japan}

\begin{abstract}
We study the propagation of a neutrino in a medium that consists of two
or more thermal backgrounds of electrons and nucleons
moving with some relative velocity,
in the presence of a static and homogeneous electromagnetic field.
We calculate the neutrino self-energy and dispersion relation using
the linear thermal Schwinger propagator, we give the formulas for the
dispersion relation and discuss general features of the results obtained,
in particular the effects of the stream contributions.
As a specific example we discuss in some detail the case of a
magnetized two-stream electron, i.e., two electron backgrounds
with a relative velocity $\vec v$ in the presence of a magnetic field.
For a neutrino propagating with momentum $\vec k$,
in the presence of the stream the neutrino dispersion relation acquires
an anisotropic contribution of the form $\hat k\cdot\vec v$ in addition
to the well known term $\hat k\cdot\vec B$, as well as
an additional contribution proportional to $\vec B\cdot\vec v$.
We consider the contribution from a nucleon stream background as an example
of other possible stream backgrounds, and comment on
possible generalizations to take into account the effects of
inhomogeneous fields. We explain why a term of the form
$\hat k\cdot(\vec v\times\vec B)$ does not appear in the dispersion
relation in the constant field case, while a term of similar form
can appear in the presence of an inhomogeneous field involving its gradient.
\end{abstract}

%\pacs{98.70.Rz; 98.70.Sa; 98.70.Vc}% PACS, the Physics and Astronomy                     
                       % Classification Scheme.
%\keywords{Suggested keywords}%Use showkeys class option if keyword
                              %display desired
\maketitle

%%%%%%%%%%%%%%%%%%%%%%%%%%%%%%%%%%%
\section{Introduction and Summary}
\label{sec:introduction}

Since the discovery of the MSW effect\cite{Wolfenstein:1977ue,Langacker:1982ih,Mikheev:1986gs},
for many years a lot of attention has been given to the calculation
of the properties of neutrinos in a matter background under various
conditions. The matter background modifies the neutrino dispersion
relations \cite{Notzold:1987ik,Pal:1989xs,Nieves:1989ez,DOlivo:1992lwg},
and also induces electromagnetic couplings that can lead
to effects in several astrophysical and/or cosmological
settings \cite{Raffelt:1996wa}.
In supernova environments the presence of the neutrino background
leads to neutrino collective oscillations \cite{Pantaleone:1992eq,Samuel:1993uw,Kostelecky:1993yt,Duan:2010bg,Mirizzi:2009td,Chakraborty:2016yeg}
that have been the subject of significant work in the context of instabilities
in supernovas\cite{Mirizzi:2009td,Chakraborty:2016yeg,Akhmedov:2016gzx}.

It is now well known that the presence of a magnetic field produces
an angular asymmetry in the neutrino dispersion relation when it propagates
in an otherwise isotropic background medium \cite{DOlivo:1989ued}. Since
many of the physical environments of interest in the contexts
mentioned include the presence of a magnetic field,
a significant amount of work has been dedicated to study the calculation
of the neutrino self-energy in the presence of a
magnetic field, or a magnetized background medium \cite{Erdas:2009zh},
and the study of the properties and propagation of neutrinos in such
media  \cite{Ioannisian:2017mqy}.

%%%%%%%%%%%%
In the previous calculations of the neutrino dispersion relation or
index of refraction in matter in the presence of a magnetic
field\cite{Lobanov:2001ar,Grigoriev:2002zr,Studenikin:2004bu,Arbuzova:2009uj}, the electron and nucleon backgrounds are
taken to be at rest since there is no other reference frame defined
in the problem at hand. In the present work we extend those
calculations by considering a medium that contains various \emph{stream}
matter backgrounds, which have a non-zero velocity relative to each
other, and including the presence of a magnetic field. The effects of moving and polarized matter
on neutrino spin/magnetic moment oscillations and $\nu_L \rightarrow
\nu_R$ conversions have been studied by several authors\cite{Lobanov:2001ar,Grigoriev:2002zr,Studenikin:2004bu,Arbuzova:2009uj}. We emphasize that our focus is different. We are concerned with
the calculation of the index of refraction or dispersion relation in
the magnetized stream media for a chiral Standard Model neutrino
state. 

In the context
of plasma physics the propagation of photons in magnetized or
unmagnetized \emph{two stream} plasma systems is a well studied
subject\cite{Shaisultanov:2011hc,Yalinewich:2010,Soto:2010,Che:2009yh,Oraevsky:2003cf}. Here
we consider the analogous problem for 
neutrinos. It is expected on general grounds that the presence of the
streams will produce corrections to both the the anisotropic and
isotropic terms in the neutrino dispersion relations, which depend on
the stream relative velocities and the magnetic field. Our goal is to
determine the corrections to the neutrino dispersion relation for a
neutrino that propagates in such magnetized \emph{stream} systems. Beyond the
intrinsic interest, the results are of practical application in
astrophysical contexts in which the asymmetric neutrino propagation is
believed to produce important effects such as the dynamics of
pulsars\cite{Kusenko:1996sr,Maruyama:2013rt} and 
supernovas\cite{Sahu:1998jh,Duan:2004nc,Gvozdev:2005}.

%%%%%%%%%%%%%%%%%%%%%%

In the previous calculations related to the propagation of neutrinos
in a matter, including the presence of a magnetic field, the electron
and nucleon backgrounds are taken to be at rest since there is no
other reference frame defined in the problem at hand. In the present
work, we consider the case in which the medium contains various \emph{stream}
matter backgrounds, which have a non-zero velocity relative to each
other, and including the presence of a magnetic field.

Before embarking on the details, we state our assumptions more precisely.
We assume that the medium contains a matter background and
a magnetic field. In the common notation, the velocity four-vector
of this background is denoted by $u^\mu$, and its reference frame
is defined by setting
\beq
\label{urestframe}
u^\mu = (1,\vec 0)\,.
\eeq
We will refer to it as the \emph{normal} background.
We assume that in that frame there is a constant magnetic field
$\vec B = B\hat b$, and in that frame we define
\beq
B^\mu = B b^\mu, \qquad b^\mu = (0, \hat b)\,.
\eeq
We can then write the corresponding EM tensor in the form
\beq
\label{F}
F_{\mu\nu} = \epsilon_{\mu\nu\alpha\beta} u^\alpha B^\beta \,,
\eeq
%
% where
%
% $\tilde P_{\mu\nu}$  was defined as $P_{\mu\nu}$ in
% Eq. (2.4) of \Ref{nuemvertexucleons},
%
%\beq
%\tilde P_{\mu\nu} = i\epsilon_{\mu\nu\alpha\beta} u^\alpha b^\beta \,.
%\eeq
%
and its dual, defined as usual by
$\tilde F_{\mu\nu} = \frac{1}{2}\epsilon_{\mu\nu\alpha\beta}F^{\alpha\beta}$,
is given by
\beq
\label{Ftilde}
\tilde F_{\mu\nu} = B_\mu u_\nu - u_\mu B_\nu\,.
\eeq
$F_{\mu\nu}$ and $\tilde F_{\mu\nu}$ are such that
\beqa
\label{FFtildeurelations}
F_{\mu\nu} u^\nu & = & 0\,,\nonumber\\
\tilde F_{\mu\nu} u^\nu & = & B_\mu\,.
\eeqa

In the present work we assume that there are additional backgrounds,
to which we refer to as the \emph{stream} backgrounds, which are superimposed
on the normal matter background having non-zero velocity relative to
the normal matter background. For definiteness, we consider only
the contributions from the electrons
and nucleons ($N = n,p$) in both the backgrounds,
and to refere them we use the symbols $s = e,N$ and $s^\prime = e^\prime,N^\prime$
respectively. We also use $f_e = e,e^\prime$ to refer
the electrons in either background, and similarly for the
nucleons $f_N = N,N^\prime$. The symbol $f$ stands for any fermion
in either background. In particular,
$u^{\mu}_f$ denotes the velocity four-vector of any of the backgrounds.

As already stated the normal background can be taken to be at rest,
so that for all the species in the normal background we set
\beq
\label{u}
u^\mu_s = u^\mu \,.
\eeq
But for the stream backgrounds
\beq
\label{uprime}
u^{\mu}_{s^\prime} = (u^0_{s^\prime}, \vec u_{s^\prime}) \,,
\eeq
in that same frame.

The main objective of the present work is the calculation of
the neutrino dispersion relations with the 
simultaneous presence of the stream background and the magnetic field.
Our work is based on the calculation of the thermal self-energy
diagrams shown in figure \ref{fig1}, using the thermal Schwinger propagator,
linearized in $B$, including only the electrons in both backgrounds,
and to the leading order $O(1/m^2_W)$ terms. The results of the
calculation are summarized in \Eqsto{bcnucleonsfinal}{bcelectronfinal}
for the self-energy, and in \Eqsto{disprelfinal}{delta}
for the corresponding dispersion relations. 
The main result is that for a neutrino propagating with momentum $\vec k$
in the presence of a stream, the neutrino dispersion relation acquires an
anisotropic contribution of the form $\hat k\cdot\vec u_{s^\prime}$
in addition to the well known term $\hat k\cdot\vec B$, 
and the standard isotropic term receives an additional contribution
proportional to $\vec B\cdot\vec u_{s^\prime}$. The term involving
$\hat k\cdot(\vec u_{s^\prime}\times\vec B)$ does not appear in the dispersion
relation, due to time-reversal invariance.

In \Section{sec:general} we summarize the general parametrization of
the self-energy, review the relevant formulas for the electron
thermal propagator and the main ingredients involved in the
calculation are given in \Section{sec:thermalpropagator}.
The formulas for the parameter coefficients that appear
in the neutrino thermal self-energy are obtained and summarized
in \Section{sec:calculations}. The calculation of the contribution
of a nucleon stream is also given there as an illustration
of possible generalizations. In \Section{sec:discussion} we discuss
and summarize the main features of the results obtained for the
neutrino dispersion relation, and comment on related work,
in particular the calculation in the case of an inhomogeneous external
field.

%%%%%%%%%%%%%%%%%%%%%%%%%%%%%%%%%%%

\section{General considerations}
\label{sec:general}

%\subsection{Self-energy}
%\label{sec:generalselfnergy}

\begin{figure}%fig1
%\vspace{-0.3cm}
{\centering
\resizebox*{0.35\textwidth}{0.3\textheight}
%\resizebox*{0.8\textwidth}{0.5\textheight}
{\includegraphics{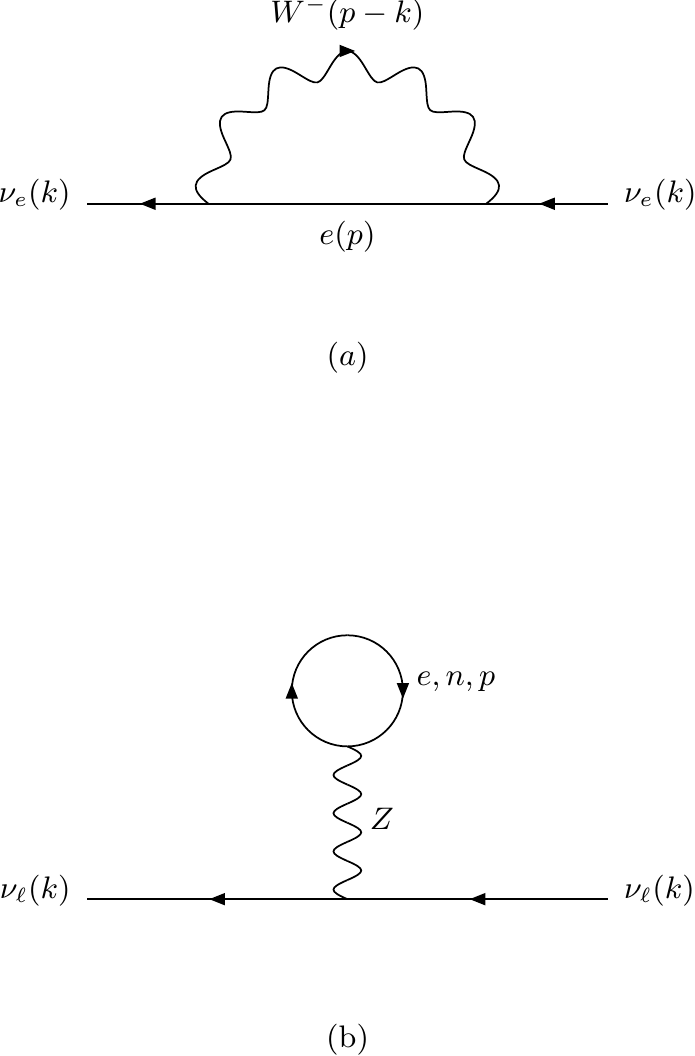}}
\par}
\caption{
The diagrams that contribute to the neutrino self-energy
  in a background of electrons and nucleons to the lowest order for a given
  neutrino flavor $\nu_\ell \; (\ell = e,\mu,\tau)$. Diagram (a) contributes
  only to the $\nu_e$ self-energy, while Diagram (b) contributes
  for the three neutrino flavors. In our calculation we consider
  two sets of these two diagrams, one set with the normal background
  ($s = e,n,p$) and another set with the stream backgrounds
  ($s^\prime = e^\prime,n^\prime,p^\prime$).
  \label{fig1}
}
\end{figure}

We denote by $\Sigma_{eff}$ the background-dependent contribution to the
neutrino self-energy, determined from the calculation of the diagrams in
figure \ref{fig1}.

%%%%%%%%%%%%%%%%%%%%%%%%%%%%%%%%%%%%%%%%%%%
%%%%%%%%%%%%%%%%%%%%%%%%%%%%%%%%%%%%%%%%%%%
%
Chirality of the neutrino interactions then imply that
\beq
\Sigma_{eff} = R\Sigma L\,,
\eeq
and the dispersion relation for a given neutrino flavor $\nu_{\ell}$
is then obtained by solving the equation
\beq
\label{efffieldeq}
\left(\lslash{k} - \Sigma\right)\psi_L = 0\,.
\eeq
In the lowest (1-loop) order each background gives a separate contribution
$\Sigma_f$ to the total self-energy.
%
%\beq
%\Sigma = Sum\; \Sigma_f\,.
%\eeq
%
In the presence of a constant electromagnetic field each term
$\Sigma_f$ is a function of $k^\mu$,
$u^\mu_f$ as well as $F^{\mu\nu}$, and its general form is
\beqa
\label{Sigmafgeneral}
\Sigma_f &=& a_f\lslash{k} +
b_f\lslash{u}_f +
c_f \tilde F^{\mu\nu}u_{f\nu}\gamma_\mu  +
d_f F^{\mu\nu}u_{f\nu}\gamma_\mu  \nonumber\\
&&+ 
g_f F^{\mu\nu} k_{\nu}\gamma_\mu +
\tilde g_f \tilde F^{\mu\nu} k_{\nu}\gamma_\mu\,.
\eeqa
In the present calculation we restrict ourselves to the contact $O(1/m^2_W)$
term of the $W$ propagator, and we do not consider the momentum dependent
terms nor its dependence on the magnetic field. To this order in $1/m^2_W$
the $a_f, g_f, \tilde g_f$ terms in \Eq{Sigmafgeneral} vanish
and we do not consider them any further. Regarding the other terms,
for our particular case in which the field is a pure $B$ field in the
rest frame of the normal background, \Eq{FFtildeurelations} implies
that $\Sigma_s$, $(s=e,n,p)$, is reduced to
\beq
\label{normalSigma}
\Sigma_s = b_s\lslash{u} + c_s\lslash{B}\,,
\eeq
which is the form used in ref. \cite{DOlivo:1989ued}. However, for the stream backgrounds,
using \Eq{Ftilde},
\beq
\label{streamSigma}
\Sigma_{s^\prime} = b_{s^\prime}\lslash{u}_{s^\prime} +
c_{s^\prime}\left[(u\cdot u_{s^\prime})\lslash{B} -
(B\cdot u_{s^\prime})\lslash{u}\right] - 
d_{s^\prime}\lslash{E}_{s^\prime}\,,
\eeq
where
\beq
E^\mu_{s^\prime} = F^{\mu\nu} u_{s^\prime\nu} =
\epsilon^{\mu\nu\alpha\beta} u_{s^\prime\nu} u_\alpha B_\beta \,.
\eeq
In the rest frame of the normal background $E^\mu_{s^\prime}$ has components
\beq
E^\mu_{s^\prime} = (0, \vec u_{s^\prime}\times\vec B)\,,
\eeq
which can be interpreted as the electric field that the stream background
particles ``see''. Thus the $d_{s^\prime}$ term represents
an electric dipole type of coupling of the stream background particles.
As we will see, the $d_{s^\prime}$ is actually not present
in our final result for $\Sigma_{s^\prime}$, which we
understand as a consequence of the fact that such couplings
require time-reversal violating effects for which there
is no source in the context of our calculation. We will discuss this
in further detail in \Section{sec:discussion}.

In summary, the contribution from each background to the self-energy
can be parametrized in the form
\beq
\label{Sigmafourcase}
\Sigma_f = b_f\lslash{u}_f + c_f \tilde F^{\mu\nu}u_{f\nu}\gamma_\mu\,.
\eeq
Therefore for the total self-energy we can write
\beq
\label{SigmaV} 
\Sigma =V^{\mu}\gamma_{\mu}\,,
% =\lslash {V}\,,
\eeq
with
\beq
\label{Vgeneral}
V^\mu = \sum_f\left(b_f u^\mu_f + c_f\tilde F^{\mu\nu} u_{f\nu}\right)\,,
\eeq
which using \Eq{Ftilde} can be expressed in the equivalent form
\beq
\label{V}
V^\mu = \sum_f\left\{b_f u^\mu_{f} +
c_f\left[(u\cdot u_f)B^\mu - (B\cdot u_f) u^\mu\right]\right\}\,.
\eeq
%
%$V^\mu$ can be expressed in the form
%
%\beq
%\label{V2}
%V^\mu = b u^\mu + c B^\mu +
%\sum_{s^\prime} b_{s^\prime} u^\mu_{s^\prime} \,,
%\eeq
%
%with
%
%\beqa
%b & = & \sum_s b_s - \sum_{s^\prime} (B\cdot u_{s^\prime})c_{s^\prime}\,,
%\nonumber\\
%c & = & \sum_s c_s + \sum_{s^\prime} (u\cdot u_{s^\prime})c_{s^\prime}\,.
%\eeqa
%

%%%%%%%%%%%%%%%%%%%%%%%%%%%%%%%%%%%

\section{Thermal propagators}
\label{sec:thermalpropagator}
\subsection{Electron propagator}
\label{sec:electronpropagator}

The internal fermion lines in the diagrams in figure \ref{fig1} stand
for the thermal fermion propagator in an external electromagnetic field,
for which we will adopt the linearized Schwinger propagator used in
ref. \cite{Chyi:1999fc,Nieves:2004qp}. 
We consider first the propagator for the
electron, for either the normal or stream background, and use the notation
$f_e = e,e^\prime$ to refer to any of them. Following that reference,
we write the Schwinger propagator (in the vacuum) in the form
\beq
\label{SF}
S^{(e)}_F = S^{(e)}_0 + S^{(e)}_B\,,
\eeq
where $S^{(e)}_0$ is the free propagator
\beq
S^{(e)}_0  =  \frac{\lslash{p} + m_e}{p^2 - m^2_e + i\epsilon}\,,
\eeq
and $S^{(e)}_B$ is the linearized $B$-dependent part of the Schwinger
propagator for the
electron\cite{note,Zhukovsky:1993kj,Borisov:1997zm,Eminov:2015uwa,Eminov:2016zqk}

%\footnote{% 
%We follow the convention that $e$ stands for the electric charge of
%the electron ($e < 0$).},
%
\beq
\label{SB}
S^{(e)}_B = \frac{eBG_e}{(p^2 - m^2_e + i\epsilon)^2}\,,
\eeq
with
\beq
\label{G}
G_e(p) = \gamma_5\left[(p\cdot b)\lslash{u} - (p\cdot u)\lslash{b} +
m_e\lslash{u}\lslash{b}\right]\,.
\eeq
Ordinarily the thermal propagator is then constructed by the
rule\cite{Elmfors:1996gy},
\beq
\label{normalthermalprop}
S^{(e)}_{11} = S^{(e)}_F - \left[S^{(e)}_F - \bar S^{(e)}_F\right]
\eta(p\cdot u)\,,
\eeq
with $\eta$ defined, as usual (see below), in terms of the
distribution function of the background electrons, and
\beq
\bar S^{(e)}_F = \gamma^0 S^{(e)\dagger}_F\gamma^0\,.
\eeq
For the calculation in this work the propagator
for each electron background ($f_e = e,e^\prime$)
is taken to be similar to \Eq{normalthermalprop},
but with $\eta(p\cdot u) \rightarrow \eta_{f_e}(p\cdot u_{f_e})$, i.e.,
\beq
\label{fthermalprop}
S^{(f_e)}_{11} = S^{(e)}_F - \left[S^{(e)}_F - \bar S^{(e)}_F\right]
\eta_{f_e}(p\cdot u_{f_e})\,,
\eeq
with $S^{(e)}_F$ as defined in \Eq{SF}. For any background fermion $f$
the function $\eta_f(p\cdot u_f)$ is given by
\beq 
\label{etaf} 
\eta_f(p\cdot u_f) = \theta(p\cdot u_f)f_f(p\cdot u_f) +
\theta(-p\cdot u_f)f_{\bar f}(-p\cdot u_f)\,, 
\eeq
with
\beqa 
f_f(x) & = & \frac{1}{e^{\beta_f(x - \mu_f)} + 1}\,, \nonumber\\[12pt]
f_{\bar f}(x) & = & \frac{1}{e^{\beta_f(x + \mu_f)} + 1} \,,
\eeqa
$\beta_f$ and $\mu_f$ being the inverse temperature and the
chemical potential of the background.
$S^{(f_e)}_{11}$ can be written in the form
\beq
\label{S11electron}
S^{(f_e)}_{11} = S^{(e)}_0 + S^{(e)}_B + S^{(f_e)}_{T} + S^{(f_e)}_{TB} \,.
\eeq
where $S^{(e)}_0$ and $S^{(e)}_B$ are the background-independent terms,
given above, while $S^{(f_e)}_{T}$ is the thermal, but $B$-independent, part
\beq
iS^{(f_e)}_{T} = -2\pi\delta(p^2 - m^2_e)\eta_{f_e}(p\cdot u_{f_e})
(\lslash{p} + m_e)\,,
\eeq
and
\beq
\label{STB}
iS^{(f_e)}_{TB} = (eB)2\pi\delta^\prime(p^2 - m^2_e)
\eta_{f_e}(p\cdot u_{f_e})G_e(p)\,,
\eeq
which is the part that is of most interest to us.
Notice that the factor $G_e(p)$ that appears here is the same for both
the normal and stream backgrounds, defined in \Eq{G}, since it refers
to the $B$-dependent part of the \emph{vacuum} Schwinger propagator.
It is useful to note that
\beq
\label{GtilfeFrel}
BG_e(p) = \tilde F^{\mu\nu} p_\nu\gamma_\mu\gamma_5 +
\frac{i}{2}m_e\tilde F^{\mu\nu}\sigma_{\mu\nu}\gamma_5\,,
\eeq
where $\tilde F_{\mu\nu}$ is given in \Eq{Ftilde} and
$\sigma_{\mu\nu} = \frac{i}{2}[\gamma_\mu,\gamma_\nu]$.

\subsection{Nucleon propagator}
\label{subsec:nucleonpropagator}

A nucleon ($N = n,p$) has an anomalous magnetic moment
coupling that also contributes to the $B$-dependent part of the
neutrino thermal self-energy. The formula analogous to \Eq{S11electron}
for the thermal Schwinger propagator for a nucleon
including the anomalous magnetic moment coupling,
was obtained in ref.~\cite{Nieves:2004qp}. Adapting that result
to our case, the thermal Schwinger propagator for a nucleon in either
the normal or stream background ($f_N = N,N^\prime)$ is
\beq
S^{(f_N)}_{11} = S^{(N)}_{0} + S^{(N)}_{B} + S^{(f_N)}_{T} + S^{(f_N)}_{TB} \,,
\eeq
where $S^{(N)}_{0}$ is the free nucleon propagator, and
\beqa
\label{nucleonpropagator}
S^{(N)}_{B} & = &
\frac{e_N BG_N(p) + \kappa_N BH_N(p)}{(p^2 - m^2_N + i\epsilon)^2}\,,
\nonumber\\
iS^{(f_N)}_{T} & = & -2\pi\delta(p^2 - m^2_N)\eta_{f_N}(p\cdot u_{f_N})
(\lslash{p} + m_N)\,,\nonumber\\
iS^{(f_N)}_{TB} & = & 2\pi\delta^\prime(p^2 - m^2_N)
\eta_{f_N}(p\cdot u_{f_N})\nonumber\\
&&\times\left[e_N BG_N(p) + \kappa_N BH_N(p)\right]\,,
\eeqa
Here we denote by $u^\mu_{f_N}$ the velocity four-vector of the
nucleon background, while $e_N$ and $\kappa_N$ stand
for the nucleon electric charge and anomalous magnetic moment,
respectively. As in the electron case above,
our working rule is that \Eq{nucleonpropagator} holds
for either a normal or stream nucleon background, with
the corresponding choice of $u^\mu_{f_N}$.
$G_N(p)$ is the same function given by \Eqs{G}{GtilfeFrel}, with
the substitution $m_e\rightarrow m_N$, while
\beq
H_N(p) = (\lslash{p} + m_N)\gamma_5\lslash{u}\lslash{b}(\lslash{p} + m_N)\,.
\eeq
In analogy with \Eq{GtilfeFrel}, here we note that $H_N(p)$ can be rewritten
in the form
\beq
\label{HtilfeFrel}
BH_N(p) = (\lslash{p} + m_N)\frac{i}{2}\tilde F^{\mu\nu}\sigma_{\mu\nu}\gamma_5
(\lslash{p} + m_N)\,.
\eeq
%
 
%%%%%%%%%%%%%%%%%%%%%%%%%%%%%%%%%%%

\section{Calculation}
\label{sec:calculations}

\subsection{$W$-diagram}

For a given background $f_e = e,e^\prime$, the $W$ diagram in  figure \ref{fig1}
gives a contribution to the neutrino thermal self-energy
\beq
\label{SigmaW}
-i\Sigma^{(W)}_{f_e} = \left(\frac{-ig}{\sqrt{2}}\right)^2\frac{i}{m^2_W}
\int\frac{d^4p}{(2\pi)^4}\gamma^\mu L iS^{(f_e)}_{11}(p)\gamma_\mu\,.
\eeq
Using \Eq{fthermalprop} and retaining only the background-dependent part,
\beq
\Sigma^{(W)}_{f_e} = \left(\Sigma^{(W)}_{f_e}\right)_{T} +
\left(\Sigma^{(W)}_{f_e}\right)_{TB}\,,
\eeq
where
\beqa
\label{SigmaWTandTB}
-i\left(\Sigma^{(W)}_{f_e}\right)_{T} & = &
\left(\frac{-ig}{\sqrt{2}}\right)^2\frac{i}{m^2_W}
\int\frac{d^4p}{(2\pi)^4}\gamma^\mu L iS^{(f_e)}_{T}(p)\gamma_\mu\,,\nonumber\\
-i\left(\Sigma^{(W)}_{f_e}\right)_{TB} & = &
\left(\frac{-ig}{\sqrt{2}}\right)^2\frac{i}{m^2_W}
\int\frac{d^4p}{(2\pi)^4}\gamma^\mu L iS^{(f_e)}_{TB}(p)\gamma_\mu\,,
\eeqa
which correspond to the $B$-independent and $B$-dependent contribution
to the neutrino thermal self-energy, respectively.
By simple Dirac algebra they can be expressed in the form
\beqa
\label{SigmaWTandTBIJ}
\left(\Sigma^{(W)}_{f_e}\right)_{T} & = &
\left(\frac{g^2}{m^2_W}\right)\lslash{I}_{f_e}\,,\nonumber\\
\left(\Sigma^{(W)}_{f_e}\right)_{TB} & = &
-\left(\frac{eg^2}{m^2_W}\right)\tilde F^{\mu\nu}
J_{f_e\nu}\gamma_\mu\,,
\eeqa
where we have used \Eq{GtilfeFrel}, and
\beqa
\label{IJdef}
I_{f\mu} & = & \int\frac{d^4p}{(2\pi)^3} \delta(p^2 - m^2_f)
\eta_f(p\cdot u_f) p_\mu\,,\nonumber\\
J_{f\mu} & = & \int\frac{d^4p}{(2\pi)^3} \delta^\prime(p^2 - m^2_f)
\eta_f(p\cdot u_f) p_\mu\,.
\eeqa
The integrals $I_{f\mu}, J_{f\mu}$ must be of the form
\beqa
\label{IJform}
I_{f\mu} & = & \tilde I_f u_{f\mu}\,,\nonumber\\
J_{f\mu} & = & \tilde J_f u_{f\mu}\,,
\eeqa
with the coefficients given by
\beqa
\label{IJtildedef}
\tilde I_f & = & \int\frac{d^4p}{(2\pi)^3} \delta(p^2 - m^2_f)
\eta_f(p\cdot u_f) p\cdot u_f\,,\nonumber\\
\tilde J_f & = & \int\frac{d^4p}{(2\pi)^3} \delta^\prime(p^2 - m^2_f)
\eta_f(p\cdot u_f) p\cdot u_f\,,
\eeqa
which can then be evaluated in any reference frame since they are scalar
integrals. A convenient one to use is the rest frame of
each background $f$. Denoting the energy and momentum of the background
particles in that reference frame by $\calE_f$ and $\vec P$,
a straightforward evaluation yields
\beqa
\label{IJtilderesult}
\tilde I_f & = & \frac{1}{4}(n_f - n_{\bar f})\,,\nonumber\\
\tilde J_f & = & -\frac{1}{2}
\int\frac{d^3 P}{(2\pi)^3 2\calE_f}\frac{d}{d\calE_f}
\left(f_f(\calE_f) - f_{\bar f}(\calE_f)\right) \,,
\eeqa
where
\beq
n_{f,\bar f} = 2\int\frac{d^3 P}{(2\pi)^3} f_{f,\bar f}(\calE_f)\,,
\eeq
and
\beq
\calE_f = \sqrt{{\vec P}^2 + m^2_f}\,.
\eeq
Therefore
\beqa
\label{SigmaWTandTBfinal}
\left(\Sigma^{(W)}_{f_e}\right)_{T} & = & b^{(W)}_f\lslash{u}_{f_e}\,,
\nonumber\\
\left(\Sigma^{(W)}_{f_e}\right)_{TB} & = &
c^{(W)}_f \tilde F^{\mu\nu} u_{f_e\nu}\gamma_\mu\,,
\eeqa
where
\beqa
b^{(W)}_{f_e} & = & \frac{g^2}{4m^2_W}(n_{f_e} - n_{\bar f_e})\,,\nonumber\\
c^{(W)}_{f_e} & = & \frac{eg^2}{2m^2_W} 
\left\{\int\frac{d^3 P}{(2\pi)^3 2\calE_e}\frac{d}{d\calE_e}
\left(f_{f_e}(\calE_e) - f_{\bar f_e}(\calE_e)\right)\right\}\,.
\eeqa

\subsection{$Z$-diagram}

For the $Z$ diagram we need the following neutral current couplings,
\beq
\label{LZ}
L_Z = -\frac{g}{2\cos\theta_W} Z^\mu \left[
\sum_{s = e,n,p}\bar s\gamma_\mu(X_s + Y_s\gamma_5)s +
\sum_\ell \bar\nu_{L\ell}\gamma_\mu\nu_{L\ell}\right] \,,
\eeq
where
\beqa
X_e & = & -\frac{1}{2} + 2\sin^2\theta_W\,,\nonumber\\
Y_e & = & \frac{1}{2}\,,
\eeqa
and %(\textbf{Verify these, and give the reference})
\beqa
\label{XYnucleons}
X_p & = & -X_e\,,\nonumber\\
X_n & = & -\frac{1}{2}\,,\nonumber\\
Y_n = -Y_p & = & \frac{1}{2}g_A\,.
\eeqa
The parameters $X_N,Y_N$ are the vector and axial vector
form factors of the nucleon neutral-current at zero momentum transfer.
In \Eq{XYnucleons}, $g_A$ stands for the normalization constant of
the axial charged vector of the nucleon, $g_A = 1.26$.

The $Z$ diagram contribution is
\beqa
&&-i\Sigma^{(Z)}_{f} = \left(\frac{-ig}{2\cos\theta_W}\right)^2
\left(\frac{i}{m^2_Z}\right) \nonumber\\
&\times& (-1)
\left\{\int\frac{d^4p}{(2\pi)^4}\mbox{Tr}\,
\gamma_\mu(X_f + Y_f\gamma_5)iS^{(f)}_{11}(p)\right\}\gamma^\mu\,,
\eeqa
and retaining only the background-dependent part,
\beq
\Sigma^{(Z)}_{f} = \left(\Sigma^{(Z)}_{f}\right)_{T} +
\left(\Sigma^{(Z)}_{f}\right)_{TB}\,,
\eeq
where
\beqa
\label{SigmafZthermal}
\left(\Sigma^{(Z)}_{f}\right)_{T} & = & -\left
  (\frac{g^2}{4m^2_W}\right )\nonumber\\
&\times&
\left\{\int\frac{d^4p}{(2\pi)^4}\mbox{Tr}\,
\gamma_\mu(X_f + Y_f\gamma_5)iS^{(f)}_{T}(p)\right\}
\gamma^\mu\,,\nonumber\\
\left(\Sigma^{(Z)}_{f}\right)_{TB} & = & -\left(
  \frac{g^2}{4m^2_W}\right )\nonumber\\
&\times&
\left\{\int\frac{d^4p}{(2\pi)^4}\mbox{Tr}\,
\gamma_\mu(X_f + Y_f\gamma_5)iS^{(f)}_{TB}(p)\right\}\gamma^\mu\,.
\eeqa
We consider the contributions from the electron and the nucleon backgrounds
separately.

\subsubsection{Electron background contribution}

In terms of the integrals $I_{f\mu}, J_{f\mu}$ defined in \Eq{IJdef},
\beqa
\label{SigmaZTandTBexpr}
\left(\Sigma^{(Z)}_{f_e}\right)_{T} & = & \left(\frac{g^2 X_e}{m^2_W}\right)
\lslash{I}_{f_e}\,,\nonumber\\
\left(\Sigma^{(Z)}_{f_e}\right)_{TB} & = & \left(\frac{eg^2 Y_e}{m^2_W}\right)
\tilde F^{\mu\nu} J_{f_e\nu}\gamma_\mu\,.
\eeqa
Comparing with \Eq{SigmaWTandTBIJ}, we then obtain
\beqa
\label{SigmaZTandTBfinal}
\left(\Sigma^{(Z)}_{f_e}\right)_{T} & = &
X_e\,\left(\Sigma^{(W)}_{f_e}\right)_{T}\,,\nonumber\\
\left(\Sigma^{(Z)}_{f_e}\right)_{TB} & = &
-Y_e\,\left(\Sigma^{(W)}_{f_e}\right)_{TB}\,,
\eeqa
with he final expressions for $\left(\Sigma^{(W)}_{f_e}\right)_{T}$
and $\left(\Sigma^{(W)}_{f_e}\right)_{TB}$ given in \Eq{SigmaWTandTBfinal}.

\subsubsection{Nucleon background contribution}
\label{subsec:nucleonstream}

We consider here the nucleon backgrounds. As with the electron case,
what interests us is the contribution to the neutrino
self-energy arising from $S^{(f_N)}_{T}$ and $S^{(f_N)}_{TB}$,
corresponding to the $B$-independent and $B$-dependent contributions
of each background ($f_N = N,N^\prime $).
Denoting them by $\left(\Sigma^{(Z)}_{f_N}\right)_{T}$ and
$\left(\Sigma^{(Z)}_{f_N}\right)_{TB}$, respectively.
From \Eq{SigmafZthermal},
% HERE
\beqa
&&-i\left(\Sigma^{(Z)}_{f_N}\right)_{X} =
\left(\frac{-ig}{2\cos\theta_W}\right)^2\left(\frac{i}{m^2_Z}\right) \nonumber\\
&\times&  (-1)
\left\{\int\frac{d^4p}{(2\pi)^4}\mbox{Tr}\,
\gamma_\mu(X_N + Y_N\gamma_5)iS^{(f_N)}_{X}(p)
\right\}\gamma^\mu\,,
\eeqa
where $X$ stands for either subscript, $T$ or $TB$.
The calculation involving $S^{(f_N)}_{T}(p)$ and the $G_N(p)$ term
of $S^{(f_N)}_{TB}(p)$ follows the steps that lead to \Eq{SigmaZTandTBexpr}.
On the other hand, using \Eq{HtilfeFrel} it follows that
\beq
\mbox{Tr}\,\gamma^\mu(X_N + Y_N\gamma_5) B H_N(p) =
-8m_N Y_N\tilde F^{\mu\nu}p_\nu\,.
\eeq
Thus we obtain
\beqa
\left(\Sigma^{(Z)}_{f_N}\right)_{T} & = & \left(\frac{g^2 X_N}{m^2_W}\right)
\lslash{I}_{f_N}\,,\nonumber\\
\left(\Sigma^{(Z)}_{f_N}\right)_{TB} & = & \left(\frac{g^2 Y_N}{m^2_W}\right)
\left(e_N + 2 m_N \kappa_N\right)\tilde F^{\mu\nu} J_{f_N\nu}\gamma_\mu\,,
\eeqa
where $I_{f_N\mu}$ and $J_{f_N\mu}$ are the integrals given by
\Eq{IJdef}, with $f = f_N$. Thus, the contribution of a
nucleon background to the neutrino self-energy is given by
\beq
\label{SigmaZnucleonsfinal}
\Sigma^{(Z)}_{f_N} = b_{f_N} \lslash{u}_{f_N} +
c_{f_N} \tilde F^{\mu\nu} u_{f_N\nu}\gamma_\mu\,,
\eeq
with
\beqa
\label{bcnucleonsfinal}
b_{f_N} & = & \frac{g^2 X_N}{4m^2_W}(n_{f_N} - n_{\bar f_N})\,,\nonumber\\
c_{f_N} & = & -\left(\frac{g^2 Y_N}{2m^2_W}\right)
\left(e_N + 2 m_N \kappa_N\right)\nonumber\\
&\times&
\left\{\int\frac{d^3 P}{(2\pi)^3 2\calE_N}\frac{d}{d\calE_N}
\left(f_N(\calE_N) - f_{\bar N}(\calE_N)\right)\right\}\,.
\eeqa
%

%%%%%%%%%%%%%%%%%%%%%%%%%%%%%%%%%%%

\subsection{Summary}
\label{subsec:selfnergysummary}

Using the results given in
\Eqss{SigmaWTandTBfinal}{SigmaZTandTBfinal}{SigmaZnucleonsfinal},
the thermal self-energy for each neutrino flavor
\beq
\label{Sigmatotalfinal}
\Sigma = \left\{
\begin{array}{ll}
  \Sigma^{(W)}_e + \Sigma^{(Z)}_e +
  \Sigma^{(W)}_{e^\prime} + \Sigma^{(Z)}_{e^\prime} & (\nu_e)\\
  \Sigma^{(Z)}_e + \Sigma^{(Z)}_{e^\prime} +
  \Sigma^{(Z)}_n + \Sigma^{(Z)}_{n^\prime} +
  \Sigma^{(Z)}_p + \Sigma^{(Z)}_{p^\prime} & (\nu_{\mu,\tau})
\end{array}
\right.
\eeq
is given by
\beq
\label{Sigmatotalfinalform}
\Sigma = \sum_{f = e,e^\prime,n,n^\prime,p,p^\prime} b_{f}\lslash{u}_f +
c_{f}\tilde F^{\mu\nu}\lslash{u}_{f\nu}\gamma_\mu\,,
\eeq
where, for $f_e = e,e^\prime$,
\beqa
\label{bcelectronfinal}
b_{f_e} & = & \frac{g^2}{4m^2_W}(n_{f_e} - n_{\bar f_e})\times\left\{
\begin{array}{ll}
1 + X_e & (\nu_e)\\
X_e & (\nu_{\mu,\tau})
\end{array}\right.\nonumber\\
c_{f_e} & = & \left(\frac{eg^2}{2m^2_W}\right)
\left\{\int\frac{d^3 P}{(2\pi)^3 2\calE_e}\frac{d}{d\calE_e}
\left(f_{f_e}(\calE_e) - f_{\bar f_e}(\calE_e)\right)\right\}\nonumber\\
&&\times\left\{
\begin{array}{ll}
1 - Y_e & (\nu_e)\\
-Y_e & (\nu_{\mu,\tau})
\end{array}\right.
\eeqa
The formulas for $b_{f_N}$ and $c_{f_N}$ ($f_N = N,N^\prime$)
are given in \Eq{bcnucleonsfinal} and they hold for any neutrino flavor.
Thus, $\Sigma$ is of the expected form discussed in
\Section{sec:general} [e.g., \Eq{Sigmafourcase}],
in particular with $d_{e^\prime} = 0$ as anticipated there,
with the coefficients $b_f,c_f$ given above in
\Eqs{bcelectronfinal}{bcnucleonsfinal}.

%%%%%%%%%%%%%%%%%%%%%%%%%%%%%%%%%%%
\section{Discussion and Conclusions}
\label{sec:discussion}

\subsection{Dispersion relations}
\label{sec:dispersionrelations}

For the purpose of determining the dispersion relations we use
the expression for $\Sigma$ in terms of $V^\mu$, \Eq{SigmaV}.
The equation for the propagating neutrino modes, \Eq{efffieldeq},
then becomes
\beq
\left(\lslash{k} - \lslash{V}\right)\psi_L = 0\,,
\eeq
and the dispersion relations are obtained by solving
\beq
k^0 - V^0 = \pm\left|\vec k - \vec V\right|\,.
\eeq
Remembering that $V^\mu$ does not depend on $k$ (to the order $1/m^2_W$
that we are considering in this work), the solutions are
$k^0 = \omega_{\pm}(\vec k)$, where
\beq
\omega_{\pm}(\vec k) = V^0 \pm \left[|\vec k| - \hat k\cdot\vec V\right]\,,
\eeq
with $\hat k$ being the unit vector along the direction of propagation.
The dispersion relation for the neutrino and the antineutrino are
identified as usual,
\beqa
\omega_\nu(\vec k) & \equiv & \omega_{+}(\vec k)\,,\nonumber\\
\omega_{\bar \nu}(\vec k) & \equiv & -\omega_{-}(-\vec k)\,,
\eeqa
which to the lowest order yield
\beq
\label{disprelfinal}
\omega_{\nu,\bar\nu}(\vec k) = |\vec k| \pm \delta\,,
\eeq
where the upper(lower) sign holds for the neutrino(antineutrino) and
\beq
\delta = V^0 - \hat k\cdot\vec V\,.
\eeq
Explicitly, using \Eq{V}, 
\beq
\delta = \sum_f \delta_f
\eeq
with
\beq
\delta_f = b_f u^0_f + c_f\vec B\cdot \vec u_f - b_f \hat k\cdot\vec u_f -
c_f u^0_f \hat k\cdot B\,.
\eeq

The fact that the dispersion relation in the presence
of a magnetic field has an anisotropic term proportional to
$\hat k\cdot\vec B$ is well known. As we have already
mentioned many of its possible effects have been studied
and more complete calculations involving higher order contributions
have been performed in the references cited. The above results
show that in the presence of a stream (with a velocity four-vector
$u^\mu_f$ relative to the normal background), the dispersion relation
acquires another anisotropic term of the form $\hat k\cdot\vec u_f$.
Furthermore, the standard isotropic (Wolfenstein) term receives
an additional contribution proportional to $\vec B\cdot\vec u_f$ that
involves the stream velocity and the magnetic field.

It has been suggested repeatedly in the literature that the
anisotropic terms in the neutrino dispersion relations
can have effects in several astrophysical environments
including pulsars \cite{Kusenko:1996sr} and the dynamics of
supernovas\cite{Sahu:1998jh,Duan:2004nc,Gvozdev:2005}.
The resonance condition
for neutrino oscillations in a magnetic field depends on $\hat k\cdot\vec B$,
and therefore is satisfied at different depths,
corresponding to different densities and temperatures. This difference
results in an asymmetry in the momentum distributions of the neutrinos.
In the presence of a stream background, the neutrino asymmetry
will depend on the relative orientation of the three vectors
$\vec k,\vec B,\vec u_f$.

As an example, let us consider specifically the two-stream electron
background. Denoting the velocity four-vector of the stream by $v^\mu$,
then
\beq
\label{delta}
\delta = b_e + c_{e^\prime}\vec B\cdot\vec v + b_{e^\prime} v^0 -
b_{e^\prime} \hat k\cdot\vec v - \left(c_e + c_{e^\prime} v^0\right)
\hat k\cdot\vec B\,.
\eeq
We wish to compare the size of the term proportional to
${\hat k}\cdot {\vec v}$ relative to ${\hat k}\cdot {\vec B}$,
thus we consider the quantity
\beq
r = \frac{b_{e^\prime}}{(c_e + c_{e^\prime} v^0)B} \,.
\eeq
For simplicity we will take $v^0 \sim 0$, and
for definiteness we will assume that the two backgrounds are described by
the classical thermal distribution functions. In that case,
\beq
\frac{df_{f}}{d\calE_f} = -\beta_f\, f_{f}\,,
\eeq
and similarly for $f_{\bar f}$, and therefore,
\beq
c_{f_e} \sim \frac{g^2}{m^2_W}\frac{\Delta N_{f_e}}{B_c}\left\{
\begin{array}{ll}
\frac{m_e}{T_{f_e}} & \mbox{(NR limit $T_{f_e} \ll m_e$)}\\
\left(\frac{m_e}{T_{f_e}}\right)^2 & \mbox{(ER limit $T_{f_e} \gg m_e$)}
\end{array}\right.
\eeq
where we have defined $\Delta N_f = n_f - n_{\bar f}$ and $B_c = m^2_e/e$.
On the other hand,
\beq
b_{e^\prime} \sim \frac{g^2}{m^2_W} \Delta N_{e^\prime}\,,
\eeq
in any case. We can consider two possibilities,
according to whether $c_e \gg c_{e^\prime}$ or the way around.
For definiteness let us consider the case $c_e \gg c_{e^\prime}$.
This situation can occur, for example, if the temperature of the normal
background is greater than the temperature of the stream. In this case 
\beq
r = \frac{b_{e^\prime}}{Bc_e} \sim
\frac{1}{B/B_c} \left(\frac{\Delta N_{e^\prime}}{\Delta N_e}\right)
\left\{
\begin{array}{ll}
\left(\frac{T_e}{m_e}\right) & \mbox{($T_e \ll m_e$)}\\
\left(\frac{T_e}{m_e}\right)^2 & \mbox{$T_e \gg m_e$)}
\end{array}\right.
\eeq
The indication is that it is possible that $r \sim 1$
for acceptable values of the parameters involved.
In other words, it is conceivable that there are environments
where the conditions are such that the asymmetries due to the
$\hat k\cdot v$ and $\hat k\cdot\vec B$ terms can be comparable.
The above formulas are based on the linear approximation in the
magnetic field and therefore are valid only for $B \ll B_c$. 

We mention that in the discussion above, in particular in writing
\Eq{delta}, we have considered a two-stream system
without explaining its physical origin, therefore in this sense
the stream velocity $\vec v$ is not specified. 
However, the results can be used in specific applications or situations
in which the stream velocity is determined and/or restricted by the
particular physical conditions of the problem, for example
if the stream velocity is due to the drift of electrons in the
$B$ field. In such a case, since the Lorentz forces makes charged particles
drift only along the $B$ axis but not in the perpendicular plane,
the results can be applied to that case as well by taking $\vec v$ to be
on the $\vec B$ axis.

Similar results are obtained in other cases as well.
To include other backgrounds we just have to add to $\delta$ the
corresponding $\delta_f$. For example, for a stream nucleon background,
\beq
\delta_{N^\prime} = b_{N^\prime} u^0_{N^\prime} +
c_{N^\prime}\vec B\cdot \vec u_{N^\prime} -
b_{N^\prime} \hat k\cdot\vec u_{N^\prime} -
c_{N^\prime} u^0_{N^\prime} \hat k\cdot B\,.
\eeq
The quantitative estimates of the effects in realistic situations
of the additional asymmetric terms that we have reported above involve stellar
astrophysics studies that are beyond the scope of the present work.
But as we have suggested they are subjects worth of further study.

\subsection{Comment on the $F^{\mu\nu}u_{f\nu}\gamma_\mu$ term}
\label{sec:dterm}

The calculations of \Section{sec:calculations} confirm explicitly
that the $d_f$ term in the general expression for the thermal self-energy
[\Eq{Sigmafgeneral}] is zero, as it was anticipated in
\Section{sec:general}. This result can be understood
by making reference to previous work \cite{Nieves:1989xg}, where the
conditions under which such dipole-type couplings may appear in the neutrino
effective Lagrangian were studied. To establish contact with that
reference, notice that the terms involving $c_f, d_f$ in
\Eq{Sigmafgeneral} are represented by the operators
\beq
O^\prime_M = c_f \tilde F^{\mu\nu} u_{f\nu}\bar\nu_L\gamma_\mu\nu_L\,,\qquad
O^\prime_E = d_f F^{\mu\nu} u_{f\nu}\bar\nu_L \gamma_\mu\nu_L\,,\qquad
\eeq
in the neutrino effective Lagrangian. The coefficients $c_f, d_f$ here
correspond to the coefficients that were denoted by $d^{\,\prime}_{M,E}$ there,
respectively (evaluated at $k = 0$). Borrowing the results of that
reference [e.g., Eqs. (14) and (16b)] the presence of $O^\prime_E$
requires time-reversal violation at some level.
% (assuming, as we do, that the hermiticity condition holds.)
Since there is no source of $T$ violation in the context of our calculation,
the $O^\prime_E$ term is not generated. On the other hand $O^\prime_M$ is
even under time-reversal but odd under $CP$, and therefore it
can be generated if the background is $CP$ asymmetric.

Here we would like to point out the following. In the presence
of non-constant fields (non-static and/or nonhomogeneous) there can be
additional terms involving the derivatives of $F_{\mu\nu}$ and/or
$\tilde F_{\mu\nu}$. For example, limiting ourselves to terms with
first derivatives, consider the following
\beq
O^{\prime\prime}_E =
h_f\left(\partial^\lambda F^{\mu\nu}\right)
u_{f\lambda} u_{f\nu}\bar\nu_L \gamma_\mu\nu_L\,.
%(u_f\cdot\partial) F^{\mu\nu} u_{f\nu}\bar\nu_L \gamma_\mu\nu_L\,.
\eeq
This term is even under $CP$ and even under time-reversal.
Therefore, it can be present in the effective Lagrangian
without implying time-reversal violation and even
if the background and the interactions are $CP$-symmetric.
This contrasts with $O^\prime_{M}$ which is $CP$-odd and therefore
does not exist if the background is $CP$-symmetric (neglecting the
$CP$ violating effects of the weak interactions). $O^{\prime\prime}_E$
can give additional anisotropic contributions to the neutrino dispersion
relation [e.g., \Eq{delta}] that are not present otherwise,
with different kinematic properties from the constant field case.
For example, in the presence of a static but inhomogeneous field,
it gives a term involving the gradient of $\hat k\cdot(\vec v\times\vec B)$.

Of course this type of term (with derivatives of the electromagnetic field)
do not appear in the approach we are using in the present work
based on the electron thermal propagator in a constant $B$ field.
Instead we have to resort to the type of approach employed
in Ref.\cite{DOlivo:1989ued}, which is based on calculating the electromagnetic vertex
first, and then taking the static limit in a suitable way to obtain
the self-energy in the (inhomogeneous) external field. We have performed
this calculation and the results are presented separately \cite{Nieves:2017rex}.

\subsection{Conclusions}

To summarize, in this work we have studied the propagation of a neutrino
in a \emph{magnetized two stream plasma system}. Specifically,
we considered a medium that consists of a \emph{normal} electron background
plus another electron \emph{stream} background that is moving as a whole
relative to the normal background. In addition, we assume that in the
rest frame of the normal background there is a constant magnetic field.

Using the thermal Schwinger propagator for the
electrons in the medium we have calculated the neutrino self-energy
in such environment, linearized in $B$ and to the leading order
$O(1/m^2_W)$ terms. The results of the calculation
are summarized in \Eqsto{bcnucleonsfinal}{bcelectronfinal}.
From the self-energy the dispersion relations were obtained
in the standard way, and the corresponding
formulas are summarized in \Eqsto{disprelfinal}{delta}.

In the presence of the stream (with velocity $\vec v$ relative
to the normal background), the dispersion relation
acquires an anisotropic term of the form $\hat k\cdot\vec v$ in
addition to the well known term of the form $\hat k\cdot\vec B$,
and the standard isotropic term receives
an additional contribution proportional to $\vec B\cdot\vec v$ that
involves the stream velocity and the magnetic field.
We explained why a term of the form $\hat k\cdot(\vec v\times\vec B)$
does not appear in the dispersion relation, due to time-reversal invariance,
and why a term  of similar kinematic form can appear in the
presence of an inhomogeneous magnetic field, involving the derivative
of the field. We have given the explicit formulas for the dispersion relations
and outlined possible generalizations, for example to include
the nucleon contribution or the case of non-homogeneous fields.
We have made simple estimates of the magnitude of the asymmetric terms
proportional to $\hat k\cdot\vec v$ and $\hat k\cdot\vec B$,
and found that they can be comparable for acceptable values of the parameters
involved.

In the context of plasma physics the propagation of photons
in \emph{two stream plasma} systems is a well studied subject.
Here we have started to carry out an analogous study for the case of neutrinos.
The present work is limited in several ways, for example by restricting
ourselves to an electron background and stream, the linear approximation
in the $B$ field, and the calculation of only the leading $O(1/m^2_W)$ terms.
However, the results reveal interesting effects that
are potentially important in several physical contexts,
such as supernova dynamics and gamma-ray bursts physics where
the effects of such systems are a major focus of current research,
and in this sense our work motivates and paves the way for
further calculations without these simplifications.

%\acknowledgments
S.S is thankful to Japan Society for the promotion of science (JSPS)
for the invitational fellowship. The work of S.S. is partially supported
by DGAPA-UNAM (M\'exico) Project No. IN110815 and PASPA-DGAPA, UNAM.

% The bibliography will probably be heavily edited during typesetting.
% We'll parse it and, using the arxiv number or the journal data, will
% query inspire, trying to verify the data (this will probalby spot
% eventual typos) and retrive the document DOI and eventual errata.
% We however suggest to always provide author, title and journal data:
% in short all the informations that clearly identify a document.

%


\begin{thebibliography}{99}
\bibitem{note} 
We follow the convention that $e$ stands for the electric
  charge of the electron ($e < 0$).
\end{thebibliography}
\end{document}